\def\stackunder#1#2{\mathrel{\mathop{#2}\limits_{#1}}}%

\documentclass[preprint,aps,showpacs,floatfix]{revtex4}
\usepackage{graphicx}

\begin{document}

\title{Scattering by Atomic Spins and Magnetoresistance in Dilute Magnetic
Semiconductors}
\author{M.~Foygel and A.~G.~Petukhov}
\affiliation{Physics Department, South Dakota School of Mines and
Technology, Rapid City, SD 57701}

\date{\today}

\begin{abstract}
We studied electrical transport in magnetic semiconductors, which is
determined by scattering of free carriers off localized magnetic moments. We
calculated the scattering time and the mobility of the majority and
minority-spin carriers with both the effects of thermal spin fluctuations
and of spatial disorder of magnetic atoms taken into account. These are
responsible for the magnetic-field dependence of electrical resistivity.
Namely, the application of the external magnetic field suppresses the
thermodynamic spin fluctuations thus promoting negative magnetoresistance.
Simultaneously, scattering off the built-in spatial fluctuations of the
atomic spin concentrations may increase with the magnetic field. The latter
effect is due to the growth of the magnitude of random local Zeeman
splittings with the magnetic field. It promotes positive magnetoresistance.
We discuss the role of the above effects on magnetoresistance of
non-degenerate semiconductors where magnetic impurities are electrically
active or neutral.
\end{abstract}
\maketitle

\section{Introduction}

\label{intro}

Dilute magnetic semiconductors (DMS) are materials of great promise in
modern technology because they combine semiconductor transport and magnetic
properties allowing reach and physically meaningful interplay between them. 
\cite{furdyna1988a,dietl1994a} Basically, there are two types of wide
bandgap semiconductor alloys in which cations are substituted by randomly
distributed magnetic atoms, such as Mn \cite{furdyna1988a,Ohno2001b}. In
some of them, such as (II,Mn)VI (Cd$_{1-x}$Mn$_{x}$Te, for instance) the
substituting magnetic impurity Mn is isoelectronic. We will call these
materials type I DMS. Here, the magnetic impurities are not electrically
active. However, they can effectively modify electronic transport \cite
{shapira1985a,shapira1990a} and magnetism \cite{dietl2001a,ferrand2001a} due
to exchange coupling of the free carriers spins to the spins of magnetic
atoms .\ As a result, scattering of the free carriers by the localized
magnetic moments determines their mobility that is substantially spin and,
therefore, magnetic-field dependent. It leads to giant magnetoresistance
(MR), positive or negative \cite{shapira1985a,shapira1990a,nam2003a}. (If
carriers are localized, the above coupling causes spin-polaron effects that
substantially affect the magnetotransport properties of type I DMS \cite
{dietl1983a,petukhov2000a}.)

More complex situation occurs in type II DMS, such as (III,Mn)V or (IV,Mn)
alloys where Mn atoms serve as acceptors. Examples are Ga$_{1-x}$Mn$_{x}$As 
\cite{matsukura1998a,Ohno1998b} and Ge$_{1-x}$Mn$_{x}$ \cite
{macdonald2005a,li2005a,jamet2006a} magnetic semiconductors. Here, contrary
to the type I DMS, atomic-spin scattering essentially involves charged
impurities a substantial fraction of which are magnetic interstitials. In
these materials, both the effects of the atomic-spin scattering and the
scattering off the charged impurities are related to each other. They cannot
be treated by means of a simple Matthiessen's rule \cite{ashcroft1976a}.

In this paper we will concentrate mainly on spin-dependent scattering which
determines mobility of free carriers in DMS taking into account the
spin-disorder effects that are intrinsic for these materials. There are two
sources of the spin-disorder effects in question: (a) the thermodynamic
fluctuations of atomic spins \cite{degennes1958a,haas1968a,kuivalainen2001a}%
, which are present even in the ordered type I magnetic materials with $x=$
1 \cite{degennes1958a,haas1968a}, and (b) the built-in spatial fluctuations
of local concentrations of the magnetic impurities \cite{michel2004a}, which
are substantial for type II DMS even in the absence of magnetic field due to
the long-range nature of \ Coulomb interaction.

The theory of spin-disorder scattering off the thermodynamic fluctuations of
the local magnetization due to atomic moments of the magnetic atoms has been
developed by P. G. de Gennes and J. Friedel for magnetic metals \cite
{degennes1958a} and for ordered magnetic semiconductors by C. Haas \cite
{haas1968a}. In particular, they showed the application of an external
magnetic field freezes out the above fluctuations thus leading to \textit{%
negative} MR. C. Michel \textit{et al} \cite{michel2004a} ignored such
effects but took into account scattering off the built-in fluctuations of
the local concentration of magnetic atoms only. They demonstrated that the
field-induced decrease in the mobility associated with these inhomogeneities
can be responsible for \textit{positive} MR in DMS. It is evident, however,
that both the above effects should be treated on an equal footage because
they have a common source - magnetic atoms. Such a treatment is especially
important because, as has been just mentioned, these spin-disorder
effects usually give competing contributions to MR of DMS. In this paper we
develop an approach that allows us to consistently tackle the problem of the
mobility of the majority and minority spins carriers by taking into account
the exchange, Coulomb, and deformation effects in scattering by the very
same magnetic atoms.

In Section~\ref{relaxation_time} we will calculate the relaxation time due
to scattering by magnetic impurities, which determines the mobility of free
carriers to be analyzed in Section~\ref{mobility}. In Section~\ref{GMR} we
will calculate the MR of different types DMS, which is in a substantial part
defined by the spin-disorder effects associated with scattering of the free
carriers off magnetic impurities.

\section{Relaxation Time}

\label{relaxation_time}

Let us consider a charge carrier in an extended state $\left| \phi _{\mathbf{%
k}}\left( \mathbf{r}\right) X\alpha \right\rangle $, where $\phi _{\mathbf{k}%
}\left( \mathbf{r}\right) =\left| \mathbf{k}\right\rangle $ is the Bloch
function of the band state of energy $\epsilon _{\mathbf{k}}$ and of wave
vector $\mathbf{k}$, $X^{\pm }$ is the electron spin up (+) or down (-)
function, and $\alpha $ is an eigenfunction of the atomic-spins Hamiltonian
with a temperature dependent probability $w_{\alpha }$ for the state $\alpha 
$ to occur. (For the sake of simplicity we will ignore a complex nature of
the angular momentum structure of the energy spectrum of such free carriers,
like holes \cite{ashcroft1976a,dietl2001a} in semiconductors, just assuming
that the carriers possess spin $s=\pm 1/2$.) For magnetic impurities
randomly located at points $\mathbf{R}_{i}$, the probability of their given
configuration

\begin{equation}
dF\left( \mathbf{R}_{1},...,\mathbf{R}_{M}\right) =\stackunder{i=1}{\Pi }%
\frac{d\mathbf{R}_{i}}{\Omega },  \label{probability}
\end{equation}
where $\Omega $ is the volume of the system. At a given temperature, the
magnetization of the system

\begin{equation}
M=xNg\mu _{B}\overline{\left\langle J_{z}\right\rangle }=M_{sat}\overline{%
\left\langle J_{z}\right\rangle }/J  \label{magnetization}
\end{equation}
is expressed in terms of the average projection of the component of the
atomic spin $\mathbf{J}_{i}$, located at point $\mathbf{R}_{i}$, along the
direction $z$ of the magnetization. Here $N$ is the concentration of the
sites in the (sub)lattice that contains magnetic atoms of the fractional
concentration $x=N_{m}/N\leq 1$, $g$ is the Lande-factor of the
magnetic-atom spin, $\mu _{B}$ is the Bohr magneton, $M_{sat}=xNg\mu _{B}J$
is the saturation magnetization; the brackets represent thermal averaging
while the bar represents the averaging over the spatial configurations of
magnetic atoms:

\begin{equation}
\overline{\left\langle J_{z}\right\rangle }=\int dF\left( \mathbf{R}_{1},...,%
\mathbf{R}_{M}\right) \sum_{\alpha }w_{\alpha }\left\langle \alpha \left|
J_{iz}\right| \alpha \right\rangle =JB_{J}\left( y\right) .  \label{J_av}
\end{equation}
Here $B_{J}\left( y\right) $ is the Brillouin function \cite{ashcroft1976a}
of the atomic spin $J$ and of the argument $y$ to be found from the
mean-field equation \cite{furdyna1988a,dietl1994a}, which depends on the
absolute temperature $T$ and external magnetic field $H$. (For the DMS well
into paramagnetic phase, $y=g\mu _{B}HJ/T$.) Throughout this paper, the
temperature $T$ is measured in the energy units ($k_{B}=1$).

We will start our consideration of a scattering time from the simplest case
of a free carrier coupled to randomly distributed non-magnetic impurities by
means of the non-exchange interaction:

\begin{equation}
U_{n-m}\left( \mathbf{r}\right) =-\sum\limits_{i}V_{n-m}\left( \mathbf{r}-%
\mathbf{R}_{i}\right) .  \label{non-exchange}
\end{equation}
The probability, per unit time, of a free-carrier transition from a state
with the wave vector $\mathbf{k}$ to a state $\mathbf{k}^{\prime }$, which
is averaged over all possible configurations of impurities, is given by:

\begin{equation}
\overline{P_{n-m}\left( \mathbf{k,k}^{\prime }\right) }=\frac{2\pi }{\hbar }%
\overline{\left| \left\langle \mathbf{k}\left| U_{n-m}\right| \mathbf{k}%
^{\prime }\right\rangle \right| ^{2}}\delta \left( \epsilon _{\mathbf{k}%
}-\epsilon _{\mathbf{k}^{\prime }}\right) .  \label{transition}
\end{equation}
Assuming the isotropy of the dispersion \ law $\epsilon _{\mathbf{k}}$ one
can express the corresponding relaxation time that appears in the Boltzmann
transport equation in terms of the above transition probability (\ref
{transition}) as follows \cite{ashcroft1976a}:

\begin{equation}
\frac{1}{\tau _{\mathbf{k}}^{n-m}}=\frac{\Omega }{\left( 2\pi \right) ^{3}}%
\int d\mathbf{k}^{\prime }\left( 1-\widehat{\mathbf{k}}\cdot \widehat{%
\mathbf{k}}^{\prime }\right) \overline{P_{n-m}\left( \mathbf{k,k}^{\prime
}\right) }.  \label{tau}
\end{equation}
Then ignoring a spatial dependence of the periodic parts of the Bloch
functions, it is easy to show that \cite{efros1988a}

\begin{equation}
\frac{1}{\tau _{\mathbf{k}}^{(n-m)}}=\frac{1}{\left( 2\pi \right) ^{2}\hbar }%
\int d\mathbf{k}^{\prime }\left( 1-\widehat{\mathbf{k}}\cdot \widehat{%
\mathbf{k}}^{\prime }\right) \Psi _{n-m}\left( \left| \mathbf{k-k}^{\prime
}\right| \right) \delta \left( \epsilon _{\mathbf{k}}-\epsilon _{\mathbf{k}%
^{\prime }}\right) ,  \label{tau1}
\end{equation}
where

\begin{equation}
\Psi _{n-m}\left( \mathbf{k}\right) =\int d\mathbf{r\exp }\left( -i\mathbf{kr%
}\right) \overline{U_{n-m}\left( \mathbf{r}\right) U_{n-m}\left( \mathbf{0}%
\right) }  \label{Fourier}
\end{equation}
is the Fourier transform of a pair correlation function of the non-exchange
part (\ref{non-exchange}) of the random impurity potential. For a simple
isotropic dispersion law $\epsilon _{\mathbf{k}}=\hbar ^{2}k^{2}/2m$ with an
effective mass $m$, Eq. (\ref{tau1}) yields

\begin{equation}
\frac{1}{\tau _{k}^{(n-m)}}=\frac{m}{4\pi \left( \hbar k\right) ^{3}}%
\int\limits_{0}^{2k}dzz^{3}\Psi _{n-m}\left( z\right) .  \label{tau2}
\end{equation}

In magnetic semiconductor, the free carrier is coupled to the randomly
distributed magnetic atoms by the following interaction:

\begin{equation}
U_{m}\left( \mathbf{r}\right) =-\sum\limits_{i}\left[ \mathbf{sJ}%
_{i}U_{ex}\left( \mathbf{r}-\mathbf{R}_{i}\right) +V\left( \mathbf{r}-%
\mathbf{R}_{i}\right) \right] ,  \label{exchange}
\end{equation}
where $U_{ex}\left( \mathbf{r}\right) \simeq \beta _{ex}\delta \left( 
\mathbf{r}\right) $ is the exchange coupling potential strongly localized
within the unit cell containing a magnetic atom ($\beta _{ex}N\simeq 1eV$) 
\cite{dietl1994a,dietl2001a}, $\mathbf{s}$ is the electron spin; $V\left( 
\mathbf{r}\right) $ is a non-exchange part of the magnetic-impurity
potential of the Coulomb and/or deformation nature. The probability per unit
time for an electron from a state with the wave vector $\mathbf{k}$ and with
spin up (+) or down (- ) to get transferred to a state with $\mathbf{k}%
^{\prime }$ and with spin up (+) or down (-) while the state of the atomic
spins undergoes transition from $\alpha $ to $\alpha ^{\prime }$ is \cite
{haas1968a}

\begin{equation}
P_{m}\left( \mathbf{k}\pm \mathbf{\alpha ,k}^{\prime }\pm \alpha ^{\prime
}\right) =\frac{2\pi }{\hbar }\left| \left\langle \phi _{\mathbf{k}}\left( 
\mathbf{r}\right) X^{\pm }\alpha \left| U_{m}\right| \phi _{\mathbf{k}%
^{\prime }}\left( \mathbf{r}\right) X^{\pm }\alpha ^{\prime }\right\rangle
\right| ^{2}\delta \left( \epsilon _{\mathbf{k}}^{\pm }+\epsilon _{\alpha
}-\epsilon _{\mathbf{k}^{\prime }}^{\pm }-\epsilon _{\alpha ^{\prime
}}\right) .  \label{probability1}
\end{equation}
Here for the simple isotropic conduction band,\bigskip 
\begin{equation}
\epsilon _{k}^{\pm }=\epsilon _{0}^{\pm }+\frac{\hbar ^{2}k^{2}}{2m}
\label{epsilon_k}
\end{equation}
and $\Delta =\epsilon _{0}^{-}-\epsilon _{0}^{+}$ is the Zeeman splitting of
the electron spin-split conduction sub-bands.

For the scattering processes that go without spin flip, thermal averaging
over the initial spin states, summation over the final spin states, and
averaging over the impurity configurations yield

\begin{eqnarray}
\sum_{\alpha ^{\prime }}\overline{\left\langle P_{m}\left( \mathbf{k+\alpha
,k}^{\prime }+\alpha ^{\prime }\right) \right\rangle } &=&\sum_{\alpha
^{\prime }}\sum_{\alpha }w_{\alpha }\int dF\left( \mathbf{R}_{1},...,\mathbf{%
R}_{M}\right) P_{m}\left( \mathbf{k+\alpha ,k}^{\prime }+\alpha ^{\prime
}\right)  \nonumber \\
&=&\frac{2\pi }{\hbar \Omega ^{2}}\int dF\left( \mathbf{R}_{1},...,\mathbf{R}%
_{M}\right) \sum_{\alpha ^{\prime }}\sum_{\alpha }w_{\alpha }\times 
\nonumber \\
&&\sum_{i}e^{i\left( \mathbf{k}^{\prime }-\mathbf{k}\right) \mathbf{R}%
_{i}}\left\{ \frac{1}{2}U_{ex}\left( \mathbf{k-k}^{\prime }\right)
\left\langle \alpha \left| J_{iz}-\left\langle J_{iz}\right\rangle \right|
\alpha ^{\prime }\right\rangle +\right.  \nonumber \\
&&\left. \left[ V\left( \mathbf{k-k}^{\prime }\right) +\frac{1}{2}%
U_{ex}\left( \mathbf{k-k}^{\prime }\right) \left\langle J_{iz}\right\rangle
\right] \delta _{\alpha \alpha ^{\prime }}\right\} \times  \nonumber \\
&&\sum_{j}e^{i\left( \mathbf{k-k}^{\prime }\right) \mathbf{R}_{j}}\left\{ 
\frac{1}{2}U_{ex}^{*}\left( \mathbf{k}^{\prime }\mathbf{-k}\right)
\left\langle \alpha ^{\prime }\left| J_{jz}-\left\langle J_{jz}\right\rangle
\right| \alpha \right\rangle +\right.  \nonumber \\
&&\left. \left[ V^{*}\left( \mathbf{k}^{\prime }-\mathbf{k}\right) +\frac{1}{%
2}U_{ex}^{*}\left( \mathbf{k}^{\prime }-\mathbf{k}\right) \left\langle
J_{jz}\right\rangle \right] \delta _{\alpha \alpha ^{\prime }}\right\}
\delta \left( \epsilon _{\mathbf{k}}^{+}+\epsilon _{\alpha }-\epsilon _{%
\mathbf{k}^{\prime }}^{+}-\epsilon _{\alpha ^{\prime }}\right)  \nonumber \\
&=&\frac{2\pi }{\hbar \Omega ^{2}}\sum_{ij}\overline{e^{i\left( \mathbf{k}%
^{\prime }-\mathbf{k}\right) \left( \mathbf{R}_{i}-\mathbf{R}_{j}\right)
}\left[ \frac{1}{4}\left| U_{ex}\left( \mathbf{k-k}^{\prime }\right) \right|
^{2}\left( \left\langle J_{iz}J_{jz}\right\rangle -\left\langle
J_{iz}\right\rangle \left\langle J_{jz}\right\rangle \right) +\right. } 
\nonumber \\
&&\left. \overline{\left| V\left( \mathbf{k-k}^{\prime }\right) +\frac{1}{2}%
U_{ex}\left( \mathbf{k-k}^{\prime }\right) \overline{\left\langle
J_{z}\right\rangle }\right| ^{2}}\right] \delta \left( \epsilon _{\mathbf{k}%
}^{+}-\epsilon _{\mathbf{k}^{\prime }}^{+}\right) ,  \label{P1_av}
\end{eqnarray}
\newline
where $U_{ex}\left( \mathbf{k}\right) $ and $V\left( \mathbf{k}\right) $ are
the Fourier transforms of the exchange and non-exchange parts of the
magnetic impurity potential. Similarly, for the scattering accompanied by
the double spin-flip processes when both the electron and magnetic atom flip
their spins simultaneously,

\begin{eqnarray}
\sum_{\alpha ^{\prime }}\overline{\left\langle P_{m}\left( \mathbf{k+\alpha
,k}^{\prime }-\alpha ^{\prime }\right) \right\rangle } &=&\sum_{\alpha
^{\prime }}\sum_{\alpha }w_{\alpha }\int dF\left( \mathbf{R}_{1},...,\mathbf{%
R}_{M}\right) P_{m}\left( \mathbf{k+\alpha ,k}^{\prime }-\alpha ^{\prime
}\right)  \nonumber \\
&=&\frac{2\pi }{\hbar \Omega ^{2}}\int dF\left( \mathbf{R}_{1},...,\mathbf{R}%
_{M}\right) \sum_{\alpha ^{\prime }}\sum_{\alpha }w_{\alpha }\times 
\nonumber \\
&&\sum_{i}e^{i\left( \mathbf{k}^{\prime }-\mathbf{k}\right) \mathbf{R}%
_{i}}\left[ \frac{1}{2}U_{ex}\left( \mathbf{k-k}^{\prime }\right)
\left\langle \alpha \left| \left( J_{ix}+J_{iy}\right) \right| \alpha
^{\prime }\right\rangle \right] \times  \nonumber \\
&&\sum_{j}e^{i\left( \mathbf{k-k}^{\prime }\right) \mathbf{R}_{j}}\left[ 
\frac{1}{2}U_{ex}^{*}\left( \mathbf{k}^{\prime }\mathbf{-k}\right)
\left\langle \alpha ^{\prime }\left| \left( J_{ix}+J_{iy}\right) \right|
\alpha \right\rangle \right] \delta \left( \epsilon _{\mathbf{k}%
}^{+}+\epsilon _{\alpha }-\epsilon _{\mathbf{k}^{\prime }}^{-}-\epsilon
_{\alpha ^{\prime }}\right)  \nonumber \\
&=&\frac{\pi }{\hbar \Omega ^{2}}\sum_{ij}\overline{e^{i\left( \mathbf{k}%
^{\prime }-\mathbf{k}\right) \left( \mathbf{R}_{i}-\mathbf{R}_{j}\right)
}\left| U_{ex}\left( \mathbf{k-k}^{\prime }\right) \right| ^{2}\left(
\left\langle J_{ix}J_{jx}\right\rangle +\left\langle
J_{iy}J_{jy}\right\rangle \right) }\times  \nonumber \\
&&\delta \left( \epsilon _{\mathbf{k}}^{+}-\epsilon _{\mathbf{k}^{\prime
}}^{-}\right)
\end{eqnarray}
Similar expressions can be derived that describe the averaged probabilities
for the (-) $\rightarrow $ (-) and (-) $\rightarrow $ (+) processes.

As a result, one can calculate, in the first Born approximation, the inverse
relaxation time for an electron with the wave vector $\mathbf{k}$ and the
spin up (+) or down (-) as follows:

\begin{eqnarray}
\frac{1}{\tau _{\mathbf{k}}^{\pm }} &=&\frac{\Omega }{\left( 2\pi \right)
^{3}}\int d\mathbf{k}^{\prime }\left( 1-\widehat{\mathbf{k}}\cdot \widehat{%
\mathbf{k}}^{\prime }\right) \sum_{\alpha ^{\prime }}\left[ \overline{%
\left\langle P_{m}\left( \mathbf{k\pm \alpha ,k}^{\prime }+\alpha ^{\prime
}\right) \right\rangle }+\overline{\left\langle P_{m}\left( \mathbf{k}\pm 
\mathbf{\alpha ,k}^{\prime }-\alpha ^{\prime }\right) \right\rangle }\right]
\nonumber \\
&=&\frac{1}{\left( 2\pi \right) ^{2}\hbar \Omega }\int d\mathbf{k}^{\prime
}\left( 1-\widehat{\mathbf{k}}\cdot \widehat{\mathbf{k}}^{\prime }\right)
\sum_{ij}\overline{e^{i\left( \mathbf{k}^{\prime }-\mathbf{k}\right) \left( 
\mathbf{R}_{i}-\mathbf{R}_{j}\right) }\left\{ \left[ \frac{1}{4}\left|
U_{ex}\left( \mathbf{k-k}^{\prime }\right) \right| ^{2}\left( \left\langle
J_{iz}J_{jz}\right\rangle -\left\langle J_{iz}\right\rangle \left\langle
J_{jz}\right\rangle \right) \right. \right. }+  \nonumber \\
&&\left. \overline{\left| V\left( \mathbf{k-k}^{\prime }\right) \pm \frac{1}{%
2}U_{ex}\left( \mathbf{k-k}^{\prime }\right) \overline{\left\langle
J_{z}\right\rangle }\right| ^{2}}\right] \delta \left( \epsilon _{\mathbf{k}%
}^{\pm }-\epsilon _{\mathbf{k}^{\prime }}^{\pm }\right) +  \nonumber \\
&&\left. \overline{\frac{1}{4}\left| U_{ex}\left( \mathbf{k-k}^{\prime
}\right) \right| ^{2}\left( \left\langle J_{ix}J_{jx}\right\rangle
+\left\langle J_{iy}J_{jy}\right\rangle \right) }\delta \left( \epsilon _{%
\mathbf{k}}^{\pm }-\epsilon _{\mathbf{k}^{\prime }}^{\mp }\right) \right\} .
\label{tau3}
\end{eqnarray}
Here the terms in the right part, which involve the atomic spins correlation
functions, are responsible for scattering off the thermodynamic fluctuations
of atomic spins. By virtue of the fluctuation-dissipation theorem\cite
{landau1980a} the first of them can be expressed in terms of the static
value of the $z$-component (longitudinal component) of the magnetic
susceptibility $\chi _{z}=\chi _{\Vert }$:

\begin{eqnarray}
\Psi _{\parallel }\left( \mathbf{k}-\mathbf{k}^{\prime }\right) &=&\frac{1}{%
4\Omega }\sum_{i,j}\overline{e^{i\left( \mathbf{k}^{\prime }-\mathbf{k}%
\right) \left( \mathbf{R}_{i}-\mathbf{R}_{j}\right) }\left| U_{ex}\left( 
\mathbf{k-k}^{\prime }\right) \right| ^{2}\left( \left\langle
J_{iz}J_{jz}\right\rangle -\left\langle J_{iz}\right\rangle \left\langle
J_{jz}\right\rangle \right) }  \nonumber \\
&=&\frac{T\left| U_{ex}\left( \mathbf{k-k}^{\prime }\right) \right| ^{2}\chi
_{\Vert }\left( \mathbf{k}-\mathbf{k}^{\prime }\right) }{4\left( g\mu
_{B}\right) ^{2}}.  \label{chi_z}
\end{eqnarray}
Similarly,

\begin{eqnarray}
\Psi _{\perp }\left( \mathbf{k}-\mathbf{k}^{\prime }\right) &=&\frac{1}{%
4\Omega }\sum_{i,j}\overline{e^{i\left( \mathbf{k}^{\prime }-\mathbf{k}%
\right) \left( \mathbf{R}_{i}-\mathbf{R}_{j}\right) }\left| U_{ex}\left( 
\mathbf{k-k}^{\prime }\right) \right| ^{2}\left\langle
J_{ix}J_{jx}\right\rangle }  \nonumber \\
&=&\frac{1}{4\Omega }\sum_{i,j}\overline{e^{i\left( \mathbf{k}^{\prime }-%
\mathbf{k}\right) \left( \mathbf{R}_{i}-\mathbf{R}_{j}\right) }\left|
U_{ex}\left( \mathbf{k-k}^{\prime }\right) \right| ^{2}\left\langle
J_{iy}J_{jy}\right\rangle }  \nonumber \\
&=&\frac{T\chi _{\bot }\left( \mathbf{k}-\mathbf{k}^{\prime }\right) \left|
U_{ex}\left( \mathbf{k-k}^{\prime }\right) \right| ^{2}}{4\left( g\mu
_{B}\right) ^{2}},
\end{eqnarray}
where $\chi _{\bot }=\chi _{x}=\chi _{y}$ are transversal components of the
magnetic susceptibility in a direction perpendicular to magnetic field. The
second term in the right part of Eq. (\ref{tau3}) represents scattering off
a random built-in potential of magnetic atoms due to the spatial
fluctuations of their local concentrations. Similarly to Eqs (\ref{tau1})
and (\ref{Fourier}) describing scattering by non-magnetic atoms, it can be
expressed in terms of the Fourier transform of the corresponding correlation
function:

\begin{equation}
\Psi _{m}\left( \left| \mathbf{k-k}^{\prime }\right| \right) =\frac{1}{%
\Omega }\sum_{i,j}\overline{e^{i\left( \mathbf{k}^{\prime }-\mathbf{k}%
\right) \left( \mathbf{R}_{i}-\mathbf{R}_{j}\right) }\left| V\left( \mathbf{%
k-k}^{\prime }\right) \pm \frac{1}{2}U_{ex}\left( \mathbf{k-k}^{\prime
}\right) \overline{\left\langle J_{z}\right\rangle }\right| ^{2}}.
\label{Psi_m}
\end{equation}
It should be mentioned here that in the case of non-degenerate
semiconductors, where the typical wave numbers of scattered free carriers
are small, one can usually ignore the dispersion of the Fourier transforms
of the correlation functions described by Eqs (\ref{Fourier}) and (\ref
{chi_z}) - (\ref{Psi_m}). The exclusion is to be made for the charged
impurities, magnetic and/or non-magnetic, when the corresponding Fourier
transforms $V_{n-m}(\mathbf{k})$ and $V(\mathbf{k})$ involved diverge at
small $k$.

\subsection{Type I DMS}

We can further simplify expression (\ref{tau3}) for the relaxation time in
the type I magnetic semiconductor compounds where the magnetic centers are
isoelectronic. In this case, the potential of such a center consists of the
exchange and deformation components, both short-range ones,

\begin{equation}
U_{m}^{\left( i\right) }\left( \mathbf{r}\right) =-\left[ \mathbf{sJ}%
_{i}U_{ex}\left( \mathbf{r}-\mathbf{R}_{i}\right) +V\left( \mathbf{r}-%
\mathbf{R}_{i}\right) \right] =-\left[ \mathbf{sJ}_{i}\beta _{ex}+\beta
_{def}\right] \delta \left( \mathbf{r}-\mathbf{R}_{i}\right) ,
\label{potential1}
\end{equation}
where $\beta _{ex}$ is the exchange coupling constant and, for ternary solid
solutions,  $\beta _{def}=N^{-1}dE_{C}/dx$ is the deformation potential
constant of the relevant band edge $E_{C}$ \cite{efros1988a,michel2004a}.
So, exactly like for any mixed semiconductor compound (see Ref. \cite
{efros1988a}), the Fourier transform of the potential correlation function

\begin{equation}
\Psi _{m}\left( k\right) =Nx\left( 1-x\right) \left| \beta _{def}\pm \frac{1%
}{2}\beta _{ex}\overline{\left\langle J_{z}\right\rangle }\right| ^{2},
\label{Psi_m1}
\end{equation}
where $\overline{\left\langle J_{z}\right\rangle }$ is given by Eq. (\ref
{J_av}). The magnetic-field dependence $\overline{\left\langle
J_{z}\right\rangle }$  follows that of the magnetization (\ref{magnetization}%
). As a result, using simple isotropic dispersion law (\ref{epsilon_k}) and
relations similar to (\ref{tau2}), Eqs (\ref{J_av}), (\ref{chi_z}) - (\ref
{potential1}) yields

\begin{eqnarray}
\frac{1}{\tau _{I}^{\pm }\left( k\right) } &=&\frac{m\beta _{ex}^{2}k}{\pi
\hbar ^{3}}\left\{ \frac{T}{\left( 2g\mu _{B}\right) ^{2}}\left[ \chi
_{\Vert }+2\chi _{\bot }F^{\pm }\left( 1\mp \frac{2m\bar{\Delta}}{\hbar
^{2}k^{2}}\right) ^{1/2}\right] +\right.   \nonumber \\
&&\left. Nx\left( 1-x\right) \left| \frac{\beta _{def}}{\beta _{ex}}\pm 
\frac{JM}{2M_{sat}}\right| ^{2}\right\} .  \label{tauI}
\end{eqnarray}
Here $F^{\pm }=1$ if the average Zeeman band splitting \cite{dietl1983a}

\begin{equation}
\bar\Delta \left( H\right) =xN\beta _{ex}JB_{J}\left( y\right) +g^{\ast }\mu
_{B}H=\frac{\beta _{ex}M\left( H\right) }{g\mu _{B}}+g^{\ast }\mu _{B}H
\label{Zeeman}
\end{equation}
is less than $\hbar ^{2}k^{2}/2m$, otherwise $F^{+}=0$ and $F^{-}=1$. (In
Eq. (\ref{Zeeman}) $g^{\ast }$ is the electron $g$-factor.)

In the above expression (\ref{tauI}) for the relaxation time of the type I
DMS, the term in the square brackets is responsible for scattering off the
thermal fluctuations of atomic spins. In particular, the term with $\chi
_{\Vert }$ describes the scattering processes that go without spin flip. The
term with $\chi _{\bot }$ takes into account the scattering accompanied by
the double spin-flip processes. For the majority-spin carriers (+), the
latter processes (+) $\rightarrow $ (-) gradually disappear with the
increase of the applied magnetic field because the energies of the
final-state subband progressively exceed that of the initial-state subband.
As a result, the corresponding transitions become energetically less
favorable 

\begin{figure}[tbph]
\bigskip
\bigskip
\includegraphics[width=.85\linewidth]{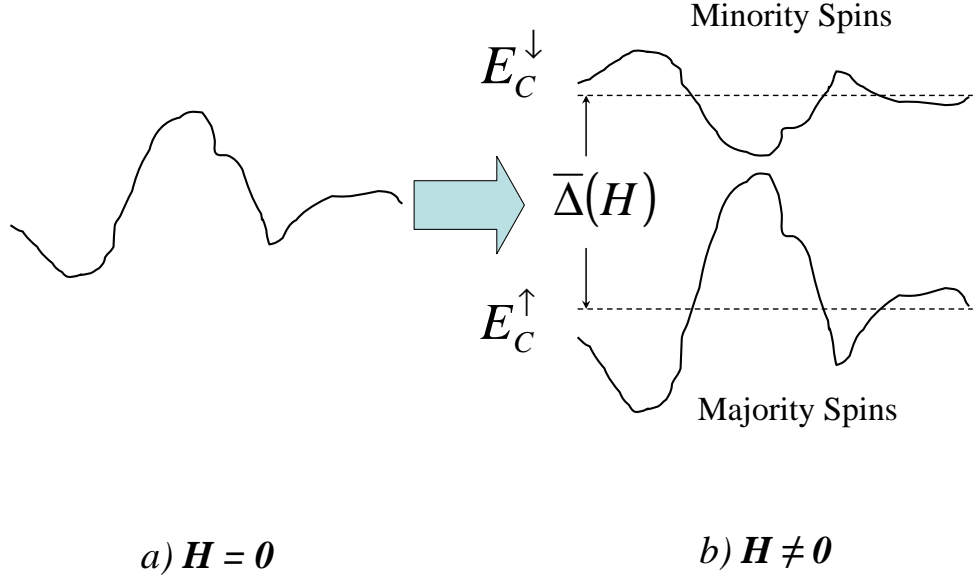} 
\vskip -100pt
\caption{(Color online.) Built-in magnetic-impurity potential in DMS in the
absence (a) and in the presence (b) of external magnetic field for the case
when $b=\beta _{def}/\beta _{ex}>1$.}
\label{potfig}
\end{figure}

The last term in the right part of Eq. (\ref{tauI}) describes scattering off
the\ spatial fluctuations of the local concentrations of magnetic atoms (see
Fig.~\ref{potfig}). For the ordered magnetic semiconductors, such as EuSe or
ErAs, where magnetic atoms form a regular sub-lattice, $x=1$, this term
disappears. Then Eq. (\ref{tauI}) coincides with one obtained by Haas \cite
{haas1968a}. Meanwhile, in the limit of the saturation magnetic fields, when
the thermal fluctuations are frozen out because both $\chi _{\Vert }\left(
H\right) =\partial M/\partial H$ and $\chi _{\bot }\left( H\right) =M/H$ \ 
\cite{haas1968a,dietl1983a} tend to zero, for $\left| \beta _{def}/\beta
_{ex}\right| >>J$ we recover the expression for the scattering time in
disordered nonmagnetic alloys \cite{efros1988a}.

\subsection{Type II DMS}

In these materials, both the magnetic impurities and the compensating
non-magnetic centers are charged, so that the spin-independent components of
their potentials are of the screened Coulomb form $V_{C}\left( r\right) $
for which the Fourier transform \cite{ashcroft1976a}

\begin{equation}
V_{C}\left( k\right) =-\frac{4\pi Ze^{2}}{\kappa \left(
k^{2}+r_{0}^{-2}\right) }.  \label{Coulomb}
\end{equation}
Here $Ze$ is the charge of a center, $\kappa $ is the dielectric constant
and $r_{0}$ is the screening length. We will use this dependence while
calculating the Fourier transforms of the relevant correlation functions (%
\ref{Psi_m}) and (\ref{Fourier}) due to magnetic and non-magnetic
impurities, respectively:

\begin{equation}
\Psi _{n-m}\left( k\right) =N_{n-m}\left[ \frac{4\pi Z_{n-m}e^{2}}{\kappa
\left( k^{2}+r_{0}^{-2}\right) }\right] ^{2},  \label{Psi_nm}
\end{equation}

\begin{equation}
\Psi _{m}\left( k\right) =N_{m}\left[ \frac{4\pi Z_{m}e^{2}}{\kappa \left(
k^{2}+r_{0}^{-2}\right) }\pm \frac{\beta _{ex}JM}{2M_{sat}}\right] ^{2},
\end{equation}
Here $Z_{i}e$ and $N_{i}$ are the charge and concentration of the magnetic ($%
m$) and non-magnetic ($n-m$) impurities. Then by employing Eq. (\ref{tau2})
it is easy to obtain the following expression for the relaxation time in the
type II DMS with large enough screening length ($kr_{0}>>1$):

\begin{eqnarray}
\frac{1}{\tau _{II}^{\pm }\left( k\right) } &=&\frac{1}{\tau _{k}^{\pm }}+%
\frac{1}{\tau _{k}^{(non-m)}}=\frac{m\beta _{ex}^{2}k}{\pi \hbar ^{3}}%
\left\{ \frac{T}{\left( 2g\mu _{B}\right) ^{2}}\left[ \chi _{\Vert }+2\chi
_{\bot }F^{\pm }\left( 1\mp \frac{2m\bar{\Delta}}{\hbar ^{2}k^{2}}\right)
^{1/2}\right] +\right.   \nonumber \\
&&\frac{N_{m}JM}{2M_{sat}}\left( \frac{JM}{2M_{sat}}\pm \frac{4\pi Z_{m}e^{2}%
}{\beta _{ex}\kappa k^{2}}\right) +\left. 2\pi N^{*}\left( \frac{e^{2}}{%
\beta _{ex}\kappa k^{2}}\right) ^{2}\ln \left( 2kr_{0}\right) \right\} ,
\label{tauII}
\end{eqnarray}
where $N^{*}=$ $N_{m}Z_{m}^{2}+N_{n-m}Z_{n-m}^{2}$ is the effective
concentration of the charged impurities. Here, like in Eq. (\ref{tauI}), the
first term in the braces is responsible for scattering off the thermodynamic
fluctuations of the atomic spins. The second and the third terms describe
the input from the built-in fluctuations of the local impurity potential. In
particular, the last term is due to scattering off charged impurities,
magnetic and non-magnetic. (For these materials we obviously have ignored
the short-range deformation potential.) It should be noted that the presence
of an ''interference'' term in the second term in the braces violates
empirical Matthiesen's rule \cite{ashcroft1976a} because the impurity
scattering processes involve the Coulomb and magnetic forces that originate
from the same atoms. It can be seen that these processes are not independent
even within the first Born approximation.

\section{Calculating Mobility}

\label{mobility}

If the relaxation time is known, one can use the standard approach to
calculate the mobilities of the majority (+) and minority (-) spin carriers 
\cite{haas1968a} 
\begin{equation}
\mu ^{\pm }=\frac{q\hbar }{3m^{2}}\frac{\sum\limits_{k}k^{2}\left( \partial
f_{k}^{\pm }/\partial \epsilon _{k}^{\pm }\right) \tau _{k}^{\pm }}{%
\sum\limits_{k}f_{k}^{\pm }},  \label{mu}
\end{equation}
where $f_{k}^{\pm }$ are the Fermi distribution functions for the spin-split
subbands and $q=\pm e$ is the carrier charge. By means of Eq. (\ref{tauI})
we find that for type I non-degenerate DMS

\begin{eqnarray}
\mu _{I}^{\pm } &=&\mu _{0}^{(I)}\int\limits_{0}^{\infty }dt\exp
(-t)t\left\{ \frac{T}{\left( 2g\mu _{B}\right) ^{2}N}\left[ \chi _{\Vert
}+2\chi _{\bot }F^{\pm }\left( 1\mp \delta /t\right) ^{1/2}\right] \right. +
\label{muI} \nonumber\\
&&\left. x\left( 1-x\right) \left| \frac{\beta _{def}}{\beta _{ex}}\pm \frac{%
JM}{2M_{sat}}\right| ^{2}\right\} ^{-1},
\end{eqnarray}
where

\begin{equation}
\mu _{0}^{(I)}\left( T\right) =\frac{2\left( 2\pi \right) ^{1/2}q\hbar ^{4}}{%
3m^{5/2}N\beta _{ex}^{2}T^{1/2}}.  \label{mu_0}
\end{equation}
Here $F^{\pm }=1$ if the dimensionless average Zeeman band splitting $\delta
=\left( \epsilon _{0}^{-}-\epsilon _{0}^{+}\right) /T=\bar{\Delta}/T\leq 1$,
otherwise $F^{+}=0$ and $F^{-}=1$. With the well known from the Hall effect
theory coefficient $\gamma _{H}=\mu _{H}/\mu =3\pi /8$ for the scattering by
isoelectronic impurities \cite{sze1981a}, in the limit of the saturating
magnetic fields and for $\left| \beta _{def}/\beta _{ex}\right| >>J$, one
can easily obtain by means of Eqs (\ref{muI}) and (\ref{mu_0}) the
expression for the Hall mobility $\mu _{H}$ in mixed non-magnetic alloys 
\cite{efros1988a}. And again, in the limiting case of the ordered magnetic
semiconductors ($x=1$) we retrieve the expression for the mobility $\mu
_{I}^{\pm }$ obtained by Haas\cite{haas1968a}.

\begin{figure}[tbhp]
\includegraphics[width=.9\linewidth]{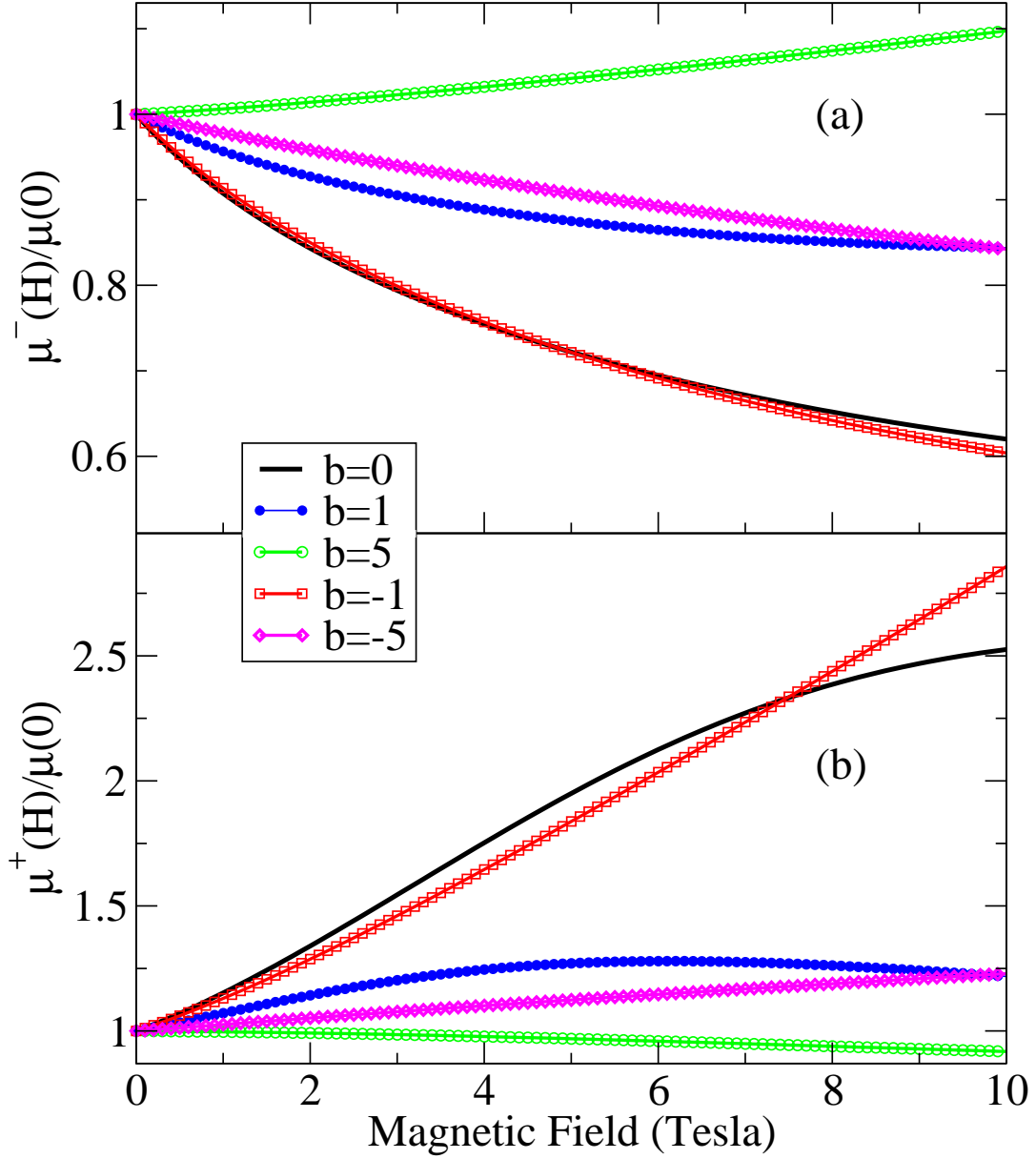}
\caption{(Color online.) Mobility of the minority (a) and majority (b) spin
carriers as a function of magnetic field for different values of the
parameter $b=\beta _{def}/\beta _{ex}$.}
\label{mufig}
\end{figure}

Fig.~\ref{mufig} shows the mobilities of spin-up and spin-down carriers in
the type I DMS calculated by means of Eq. (\ref{muI}) for different values
of parameter $b=\beta _{def}/\beta _{ex}$. We used $N\left| \beta
_{ex}\right| =1eV$ \cite{dietl2001a,ferrand2001a}, $J=5/2,$ $x=0.02$, and $%
T=50K$ in our calculations. It can be seen that, as a rule, the mobility of
spin down carriers drops while that of the spin up carriers rises with the
application of the magnetic field. The latter (b) is dominated by scattering
off the thermodynamic fluctuations of the atomic spins, which is suppressed
by magnetic field. Simultaneously, with the application of magnetic field,
the former (a) is more and more determined by scattering off the built-in
fluctuations of the local concentration of magnetic atoms because the input
from the double spin-flip (+) $\Rightarrow $(-) processes of scattering by
the thermal fluctuations are substantially reduced for these processes are
accompanied by the absorption of the increasingly greater amounts of thermal
energy. Contrary to that, the spin-up mobility drops and spin-down mobility
rises with magnetic field only if the deformation potential constant $\beta
_{def}$ is large and have the same sign as the exchange coupling constant $%
\beta _{ex}$. Then in those both cases the zero-magnetic-field mobility is
governed by scattering off the build-in fluctuations of the deformation
potential of magnetic impurities (Fig.~\ref{potfig}(a)). The application of
the magnetic field increases the amplitude of the impurity potential
fluctuations for the majority-spin carriers and decreases the above
amplitude for the minority spin carriers (Fig.~\ref{potfig}(b)), which
explains the calculated dependencies.

For the type II non-degenerate DMS with large enough screening length ($%
8mTr_{0}^{2}/\hbar ^{2}>>1$), Eqs (\ref{tauII}) and (\ref{mu}) yield

\begin{eqnarray}
\mu _{II}^{\pm } &=&\mu _{0}^{(II)}\int\limits_{0}^{\infty }dt\exp
(-t)t\left\{ \frac{T}{\left( 2g\mu _{B}\right) ^{2}N_{m}}\left[ \chi _{\Vert
}+2\chi _{\bot }F^{\pm }\left( 1\mp \delta /t\right) ^{1/2}\right] \right. +
\nonumber \\
&&\left. \frac{JM}{2M_{sat}}\left( \frac{JM}{2M_{sat}}\pm \frac{Z_{m}T_{0}}{%
Tt}\right) +\frac{N^{*}}{2\pi N_{m}}\left( \frac{T_{0}}{2Tt}\right) ^{2}\ln
\left( 8mTtr_{0}^{2}/\hbar ^{2}\right) \right\} ^{-1}.  \label{muII}
\end{eqnarray}
Here $\mu _{0}^{(II)}\sim T^{-1/2}$ is given by Eq. (\ref{mu_0}) where $N$
is to be changed for the concentration of magnetic impurities $N_{m}$. In 
Eq. (\ref{muII}), we introduced parameter $T_{0}=2\pi e^{2}\hbar ^{2}/(\beta
_{ex}\kappa m)$, which is of the order of 1 eV, so that at practically all
temperatures $T<<T_{0}$. Therefore, the last term in the expression in the
braces in the right part of Eq. (\ref{muII}) is much larger than the
previous two. As a result, the mobility here is dominated by scattering off
charged impurities and is approximately described by \cite
{conwell1950a,sze1981a}

\begin{equation}
\mu _{II}^{\pm }\simeq \mu _{CW}\left( T\right) =\frac{2^{7/2}\kappa
^{2}T^{3/2}}{N^{*}q^{3}m^{1/2}\ln \left( 24mTr_{0}^{2}/\hbar ^{2}\right) }.
\label{mu_BH}
\end{equation}
Eq. (\ref{muII}) allows someone to easily calculate small
magnetic-field-dependent corrections to the Conwell-Weisscopf expression (%
\ref{mu_BH}).

\section{Application to GMR}

\label{GMR}

An applied magnetic field changes the conductivity of DMS 
\begin{equation}
\sigma =q\left( \mu ^{+}n^{+}+\mu ^{-}n^{-}\right)  \label{conductivity}
\end{equation}
because it affects both the mobility $\mu ^{\pm }$ and the concentration of
the majority and minority spin carriers, which for non-degenerated DMS is
equal to

\begin{equation}
n^{\pm }=\frac{1}{2}N_{c}\exp \left( -\frac{E_{C}-\Gamma ^{\pm }-F\mp \bar{%
\Delta}/2}{T}\right) .  \label{concentration}
\end{equation}
Here $F$ is the Fermi energy, $N_{c}$ is the density of the conduction band
edge $E_{C\text{ }}$states and

\begin{equation}
\Gamma ^{\pm }=\frac{x\left( 1-x\right) m}{2\pi N\hbar ^{2}a}\left( \beta
_{def}\pm \beta _{ex}\frac{JM}{2M_{sat}}\right) ^{2}  \label{Gamma}
\end{equation}
is the shift of the bottom of the spin-split sub-bands, which is caused by
the renormalization of the energy spectrum by the fluctuating short-range
electronic potential in mixed ternary compounds including type I DMS \cite
{efros1988a,michel2004a}. In Eq. (\ref{Gamma}), $a$ is the lattice constant.

Let us, for the sake of simplicity, assume that the total concentration of
the free carriers, $n=n^{+}+n^{-}$, does not depend on magnetic-field. It is
usually correct when the concentration is determined by shallow non-magnetic
impurities for which the ionization energy does not depend on magnetic field 
\cite{haas1968a}. Then the MR can be calculated as

\begin{eqnarray}
\frac{\Delta \rho \left( H\right) }{\rho \left( 0\right) } &=&\frac{\sigma
\left( 0\right) }{\sigma \left( H\right) }-1=\left[ \frac{\mu ^{+}\left(
H\right) /\mu (0)}{1+\exp \left( -\left( \bar{\Delta}\left( H\right) +\gamma
(H)\right) /T\right) }+\right.   \nonumber \\
&&\left. \frac{\mu ^{-}\left( H\right) /\mu (0)}{1+\exp \left( \left( \bar{%
\Delta}\left( H\right) +\gamma (H)\right) /T\right) }\right] ^{-1}-1,
\label{MR}
\end{eqnarray}
where $\mu ^{\pm }\left( H\right) $ is given by Eq. (\ref{muI}), $\bar{\Delta%
}\left( H\right) $ is given by Eq. (\ref{Zeeman}), and $\gamma (H)=\Gamma
^{+}\left( H\right) -\Gamma ^{-}\left( H\right) $.

\begin{figure}[tbph]
\includegraphics[width=.9\linewidth]{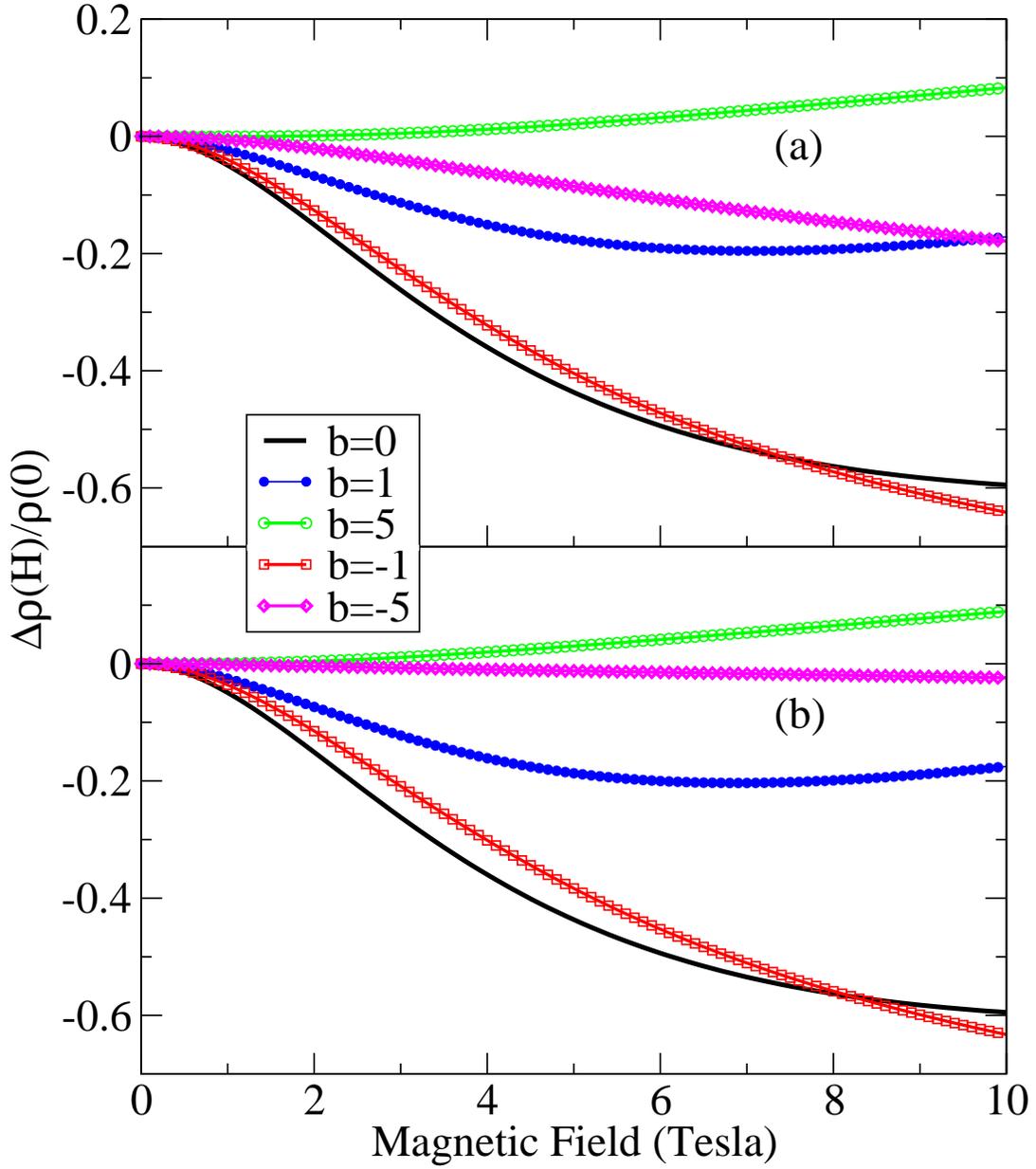}
\caption{(Color online.) Magnetoresistance of DMR for different values of
the parameter $b=\beta _{def}/\beta _{ex}$ with (b) and without (a)
band-edge shift taken into account.}
\label{mrfig}
\end{figure}

Fig.~\ref{mrfig} shows the magnetoresistance of DMS calculated by means of
Eq (\ref{MR}) with the same set of parameters we used to generate Fig.~\ref
{mufig}. 
The first set of graphs (a) represents the MR calculated without the shifts (%
\ref{Gamma}) of the spin-split band edges taken into account ($\gamma (H)=0$%
), whereas the second set (b) takes this effect into consideration. (Here we
used $m=0.1m_{e}$ and $N=4/a^{3}$ \cite{dietl2001a,ferrand2001a} in our
calculations). One can see that MR is predominantly governed by the
magnetic-field dependence of the mobility of the majority carriers, which
itself is dominated by scattering off the thermodynamic fluctuations of
atomic spins. As a result, the MR is usually negative because both the
magnitude of these fluctuations and related scattering decrease with the
application of magnetic field. For the model under discussion the MR becomes
positive only if the deformation potential constant $\beta _{def}$ is large
and has the same sign as the exchange coupling constant $\beta _{ex}$ ($%
b=\beta _{def}/\beta _{ex}=5$). In that case, the zero-magnetic field
mobility is dominated by scattering off the large build-in fluctuations of
the deformation potential of magnetic impurities. The application of the
magnetic field increases the amplitude of these fluctuations for the
majority-spin carriers thus leading to positive MR. Taking into account the
renormalization of the band edges does not substantially change the MR with
an insignificant exclusion for the case of $b=\beta _{def}/\beta _{ex}=-5$
when the MR becomes slightly less negative.

In the recent publication \cite{li2007a} the authors analyzed the observed
MR of $Mn:Ge$ DMS by using a modified version of the atomic spin scattering
model under consideration. In the paramagnetic phase, these materials reveal
large positive MR. It has been demonstrated \cite{li2007a} that this
phenomenon is related to superparamagnetic nature of magnetic clusters with
enhanced concentration of $Mn$ atoms. In addition, antiferromagnetic
coupling between magnetic atoms in the superparamagnetic clusters has been
taken into account, which, together with an anisotropy of the magnetic
susceptibility, is shown to promote positive MR.

In summary, we analyzed spin-dependent electrical conductivity in DMS where
free carriers are scattered off randomly distributed magnetic impurities.
The mobility of the minority and majority spin carriers is shown to be
governed by competing impurity spin-disorder effects caused by (a)
thermodynamic fluctuations of the atomic spins and (b) random impurity
potential generated by the spacially fluctuating concentration of the
magnetic atoms. The former effect usually dominates the mobility of the
majority-spin carriers. It is quenched by external magnetic field leading to
giant negative magnetoresistance. Depending on material parameters, the
spin-dependent scattering rate may be enhanced due the growing fluctuations
of local Zeeman splittings of the expanded electronic states leading to
positive magnetoresistance in such DMS.

This work was supported by ONR Grant N00014-06-1-0616.

\bibliographystyle{/home/andre/tex/bibtex/apsrev}
\bibliography{/home/andre/tex/bibtex/all}

\end{document}